\documentclass[%
reprint,
superscriptaddress,
showpacs,
amsmath,amssymb,
prc,
floatfix, ]%
{revtex4-1}

\usepackage{color}
\usepackage{CJK}
\usepackage{longtable}
\usepackage{graphicx}
\usepackage{dcolumn}
\usepackage{bm}
\usepackage[dvipdfm,bookmarks=true,colorlinks,%
            citecolor=blue,linkcolor=blue,hypertex, %
            breaklinks=true]{hyperref}

\begin{document}

\begin{CJK*}{GBK}{}

\title{Configuration and bandhead spin assignments for the 2-quasiparticle
       rotational bands in the neutron-rich nuclei $^{154, 156}$Pm}

\author{Shuo-Yi Liu}
 \affiliation{Mathematics and Physics Department,
              North China Electric Power University, Beijing 102206, China}
\author{Miao Huang}
 \affiliation{Mathematics and Physics Department,
              North China Electric Power University, Beijing 102206, China}
\author{Zhen-Hua Zhang}
 \email{zhzhang@ncepu.edu.cn}
 \affiliation{Mathematics and Physics Department,
              North China Electric Power University, Beijing 102206, China}
 \affiliation{Department of Physics and Astronomy,
               Mississippi State University, Mississippi 39762, USA}

\date{\today}

\begin{abstract}
The rotational bands in the neutron-rich nuclei $^{153-157}$Pm
are investigated by a particle-number conserving method.
The kinematic moments of inertia for the 1-quasiparticle bands in
odd-$A$ Pm isotopes $^{153, 155, 157}$Pm are reproduced quite well by the present calculation.
By comparison between the experimental and calculated moments of inertia for
the three 2-quasiparticle bands in the odd-odd nuclei $^{154, 156}$Pm,
their configurations and bandhead spins have been assigned properly.
For the 2-quasiparticle band in $^{154}$Pm, the configuration is assigned
as $\pi5/2^-[532]\otimes \nu3/2^-[521]$ ($K^\pi=4^+$)
with the bandhead spin $I_0=4\hbar$.
In $^{156}$Pm, the same configuration and bandhead spin assignments have been made
for the 2-quasiparticle band with lower excitation energy.
The configuration $\pi5/2^+[413]\otimes \nu5/2^+[642]$ ($K^\pi=5^+$)
with the bandhead spin $I_0=5\hbar$ is assigned for that with higher excitation energy.
\end{abstract}

\maketitle
\end{CJK*}

\section{Introduction}{\label{Sec:Introduction}}

Spectroscopic investigation of the neutron-rich nuclei around $A\approx 150-160$
mass region is quite challenging due to the fact that it is
very hard to find an appropriate combination of  projectile and target to produce
these nuclei in the fusion-evaporation reaction with sufficiently large cross section.
Fortunately, the isomeric and high-spin states of these neutron-rich nuclei can be produced
with high efficiency in the spontaneous fission of the actinide
nuclei~\cite{Hamilton1995_PPNP35-635, Hamilton1997_PPNP38-273},
by which various high-$K$ isomers and high-spin rotational bands for the neutron-rich
nuclei in this mass region have been established up to now.
These data can provide detailed information on the nuclear phenomena
such as $K$-isomerism, quantum phase transition, octupole correlations,
thus providing a benchmark for various available nuclear
models~\cite{Andersson1976_NPA268-205, Bengtsson1979_NPA327-139,
Nazarewicz1985_NPA435-397, Dobaczewski1997_CPC102-166,
Afanasjev1996_NPA608-107, Hara1995_IJMPE4-637}.

As for the Pm ($Z=61$) isotopes, a lot of efforts have been put on
searching for the parity doublet bands in order to investigate
the reflection-asymmetric shape in this transitional mass region.
Up to now, parity doublet bands in Pm isotopes have been observed in
$^{147}$Pm~\cite{Urban1995_NPA587-541-561},
$^{149}$Pm~\cite{Jones1996_NPA609-201}, and
$^{151}$Pm~\cite{Vermeer1990_PRC42-R1183, Urban1990_PLB247-238}.
Recently, high-spin structures in the neutron-rich $^{152-158}$Pm isotopes,
including both odd-$A$ and odd-odd nuclei,
have been observed experimentally by the spontaneous fission
of the actinide nuclei~\cite{Bhattacharyya2018_PRC98-044316}.
Compared with pervious experiments~\cite{Daniels1971_PRC4-919,
Dauria1971_NPA178-172, Karlewski1985_ZPA322-177, Greenwood1987_PRC35-1965,
Taniguchi1996_JPSJ65-3824, Shibata2007_EPJA31-171, Hwang2009_PRC80-037304},
these rotational bands either have been obtained for the first time,
or have been extended considerably to higher spins.
Unfortunately, these experimental data do not show any evidence of
existing of octupole deformation in these Pm isotopes with neutron number $N>90$.
In Ref.~\cite{Bhattacharyya2018_PRC98-044316}, the configurations for the
1-quasiparticle (1-qp) rotational bands in odd-$A$ Pm isotopes have been
assigned properly by the cranked relativistic Hartree-Bogoliubov calculations.
Compared to odd-$A$ nuclei, the structure of odd-odd nuclei is more complicated
due to the contributions from both valence neutrons and protons.
Note that in Refs.~\cite{Sood2011_PRC83-027303, Sood2011_PRC83-057302, Sood2012_EPJA48-1},
some configurations for the 2-qp isomeric states in $^{154, 156}$Pm have already been
suggested by the quasiparticle-rotor model.
However, due to the lack of firm spin and parity assignments, reasonable configurations
have not been assigned for these 2-qp rotational bands in odd-odd Pm isotopes
in Ref.~\cite{Bhattacharyya2018_PRC98-044316}.
In the present work, the cranked shell model (CSM) with
pairing correlations treated by a particle-number conserving
(PNC) method~\cite{Zeng1983_NPA405-1, Zeng1994_PRC50-1388}
will be used for investigating the  rotational bands in these Pm isotopes,
and the configuration and bandhead spin assignments will be made for
the three 2-qp bands observed in $^{154, 156}$Pm.
Note that PNC-CSM has already been used for the systematic investigation
of the high-$K$ isomers and high-spin rotational bands in the neighboring neutron
rich Nd ($Z=60$), Sm ($Z=62$) and Gd ($Z=64$)
isotopes~\cite{Zhang2018_PRC98-034304, He2018_PRC98-064314},
in which the experimental data are reproduced quite well.
Therefore, the calculations for Pm isotopes by PNC-CSM should be quite reliable.

Different from the conventional Bardeen-Cooper-Schrieffer
or Hartree-Fock-Bogoliubov method,
in the PNC method, the pairing Hamiltonian is diagonalized directly
in a properly truncated Fock-space~\cite{Wu1989_PRC39-666}.
Therefore, the particle-number is totally conserved
and the Pauli blocking effects are treated exactly,
and it is very suitable for the investigation of the multi-qp rotational bands.
Note that the PNC method has also been transplanted in
the total-Routhian-surface method~\cite{Fu2013_PRC87-044319},
and both relativistic~\cite{Meng2006_FPC1-38, Shi2018_PRC97-034317}
and non-relativistic mean field models~\cite{Liang2015_PRC92-064325}.
Similar exact particle-number conserving approaches can be found in
Refs.~\cite{Richardson1964_NP52-221, Pan1998_PLB422-1, Volya2001_PLB509-37,
Jia2013_PRC88-044303, Jia2013_PRC88-064321, Chen2014_PRC89-014321}.

This paper is organized as follows.
The theoretical framework of PNC-CSM is presented briefly in Sec.~\ref{Sec:PNC-CSM}.
In Sec.~\ref{Sec:Results}, the MOIs of the rotational bands in Pm isotopes
are calculated and compared with the data.
The configuration and bandhead spin assignments for the three 2-qp bands
in odd-odd nuclei $^{154,156}$Pm are made.
A brief summary is given in Sec.~\ref{Sec:Summary}.

\section{Theoretical framework}{\label{Sec:PNC-CSM}}

The cranked shell model Hamiltonian with pairing correlations
can be written as
\begin{eqnarray}
 H_\mathrm{CSM}
 & = &
 H_0 + H_\mathrm{P}
 = H_{\rm Nil}-\omega J_x + H_\mathrm{P}
 \ ,
 \label{eq:H_CSM}
\end{eqnarray}
where $H_{\rm Nil}$ is the Nilsson Hamiltonian~\cite{Nilsson1969_NPA131-1},
$-\omega J_x$ is the Coriolis interaction with
the cranking frequency $\omega$ about the $x$ axis.
$H_{\rm P} = H_{\rm P}(0) + H_{\rm P}(2)$ is the pairing Hamiltonian
with monopole and quadrupole pairing interaction,
\begin{eqnarray}
 H_{\rm P}(0)
 & = &
  -G_{0} \sum_{\xi\eta} a^\dag_{\xi} a^\dag_{\bar{\xi}}
                        a_{\bar{\eta}} a_{\eta}
  \ ,
 \\
 H_{\rm P}(2)
 & = &
  -G_{2} \sum_{\xi\eta} q_{2}(\xi)q_{2}(\eta)
                        a^\dag_{\xi} a^\dag_{\bar{\xi}}
                        a_{\bar{\eta}} a_{\eta}
  \ ,
\end{eqnarray}
where $\bar{\xi}$ ($\bar{\eta}$) is the time-reversal state of  $\xi$ ($\eta$),
$q_{2}(\xi) = \sqrt{{16\pi}/{5}} \langle \xi |r^{2}Y_{20} | \xi \rangle$
is the diagonal element of the stretched quadrupole operator,
$G_0$ and $G_2$ are the effective monopole and quadrupole pairing strengths.

In the PNC method, the pairing Hamiltonian $H_{\rm P}$ is diagonalized
directly in a sufficiently large cranked many-particle configuration
(CMPC, an eigenstate of the one-body Hamiltonian $H_0$) space~\cite{Zeng1983_NPA405-1}.
Instead of the traditional single-particle level truncation used in shell-model
calculation, a CMPC truncation is adopted, which can make the PNC calculation
both workable and accurate~\cite{Wu1989_PRC39-666,Molique1997_PRC56-1795}.
For the investigation of rare-earth nuclei,
usually a CMPC space with the dimension of 1000 is enough.
The eigenstates of $H_\mathrm{CSM}$ can be obtained
by diagonalization in the truncated CMPC space
\begin{equation}
 |\Psi\rangle = \sum_{i} C_i \left| i \right\rangle \ ,
\end{equation}
where $| i \rangle$ is a CMPC and $C_i$ is the expanding coefficient.

The angular momentum alignment for the state
$| \Psi \rangle$ can be written as
\begin{equation}
\langle \Psi | J_x | \Psi \rangle = \sum_i C_i^2 \langle i | J_x | i
\rangle + 2\sum_{i<j}C_i C_j \langle i | J_x | j \rangle \ ,
\end{equation}
and the kinematic moment of inertia (MOI) is
\begin{equation}
J^{(1)}=\frac{1}{\omega} \langle\Psi | J_x | \Psi \rangle \ .
\end{equation}

The experimental MOI and rotational frequency
for one rotational band can be extracted by
\begin{eqnarray}\label{eq:exp-moi}
\frac{J^{(1)}(I)}{\hbar^2}&=&\frac{2I+1}{E_{\gamma}(I+1\rightarrow I-1)} \ , \nonumber \\
\hbar\omega(I)&=&\frac{E_{\gamma}(I+1\rightarrow I-1)}{I_{x}(I+1)-I_{x}(I-1)} \ ,
\end{eqnarray}
separately for each signature sequence ($\alpha = I$ mod 2),
where $I_{x}(I)=\sqrt{(I+1/2)^{2}-K^{2}}$, and $K$ is the projection of
the total angular momentum onto the symmetry $z$ axis.

\section{Results and discussion}{\label{Sec:Results}}

\begin{table}[h]
 \centering
 \caption{\label{tab:def} Deformation parameters ($\varepsilon_2$, $\varepsilon_4$)
  for Pm isotopes adopted in the present PNC-CSM calculation,
  which are taken from Ref.~\cite{Moeller1995_ADNDT59-185}.}
\begin{tabular*}{1.0\columnwidth}{c@{\extracolsep{\fill}}ccccc}
\hline
\hline
                & $^{153}$Pm & $^{154}$Pm & $^{155}$Pm & $^{156}$Pm & $^{157}$Pm \\
\hline
$\varepsilon_2$ & 0.250      & 0.250      & 0.258      & 0.258      & 0.267      \\
$\varepsilon_4$ & -0.073     & -0.067     & -0.060     & -0.060     & -0.047     \\
\hline
\hline
\end{tabular*}
\end{table}

In the present calculation for Pm isotopes, the parameters in the PNC-CSM
are taken the same as our previous investigation for the neighboring
Nd and Sm isotopes~\cite{Zhang2018_PRC98-034304}.
Here we show them again briefly for convenience.
The deformation parameters ($\varepsilon_2$, $\varepsilon_4$) are taken from
Ref.~\cite{Moeller1995_ADNDT59-185} (c.f., Table~\ref{tab:def}) and
the Nilsson parameters ($\kappa$ and $\mu$) are taken as
the traditional values~\cite{Nilsson1969_NPA131-1}.
In addition, the neutron orbital $\nu5/2^+[642]$ is shifted upwards
by $0.07\hbar\omega_0$ for all Pm isotopes
to reproduce the experimental single-particle level sequence.
The CMPC space is constructed in proton $N=4, 5$ major shells
and neutron $N=5, 6$ major shells, respectively.
The CMPC truncation energies are about 0.85$\hbar\omega_0$ for both protons and neutrons.
The dimensions of the CMPC space are 1000 for both protons and neutrons.
For all Pm isotopes, the monopole and quadrupole pairing strengths are chosen as
$G_{\rm 0p}=0.25$~MeV and $G_{\rm 2p}=0.01$~MeVfm$^{-4}$ for protons,
$G_{\rm 0n}=0.30$~MeV and $G_{\rm 2n}=0.02$~MeVfm$^{-4}$ for neutrons.
Note that the paring strengths in Ref.~\cite{Zhang2018_PRC98-034304}
are determined by the odd-even differences in nuclear binding energies
of Nd and Sm isotopes [see Fig. 1 in Ref.~\cite{Zhang2018_PRC98-034304}].
Since these Pm isotopes are the neighbors of the Nd and Sm nuclei
in Ref.~\cite{Zhang2018_PRC98-034304}, their pairing strengths should be similar.

\begin{figure}[h]
\includegraphics[width=0.95\columnwidth]{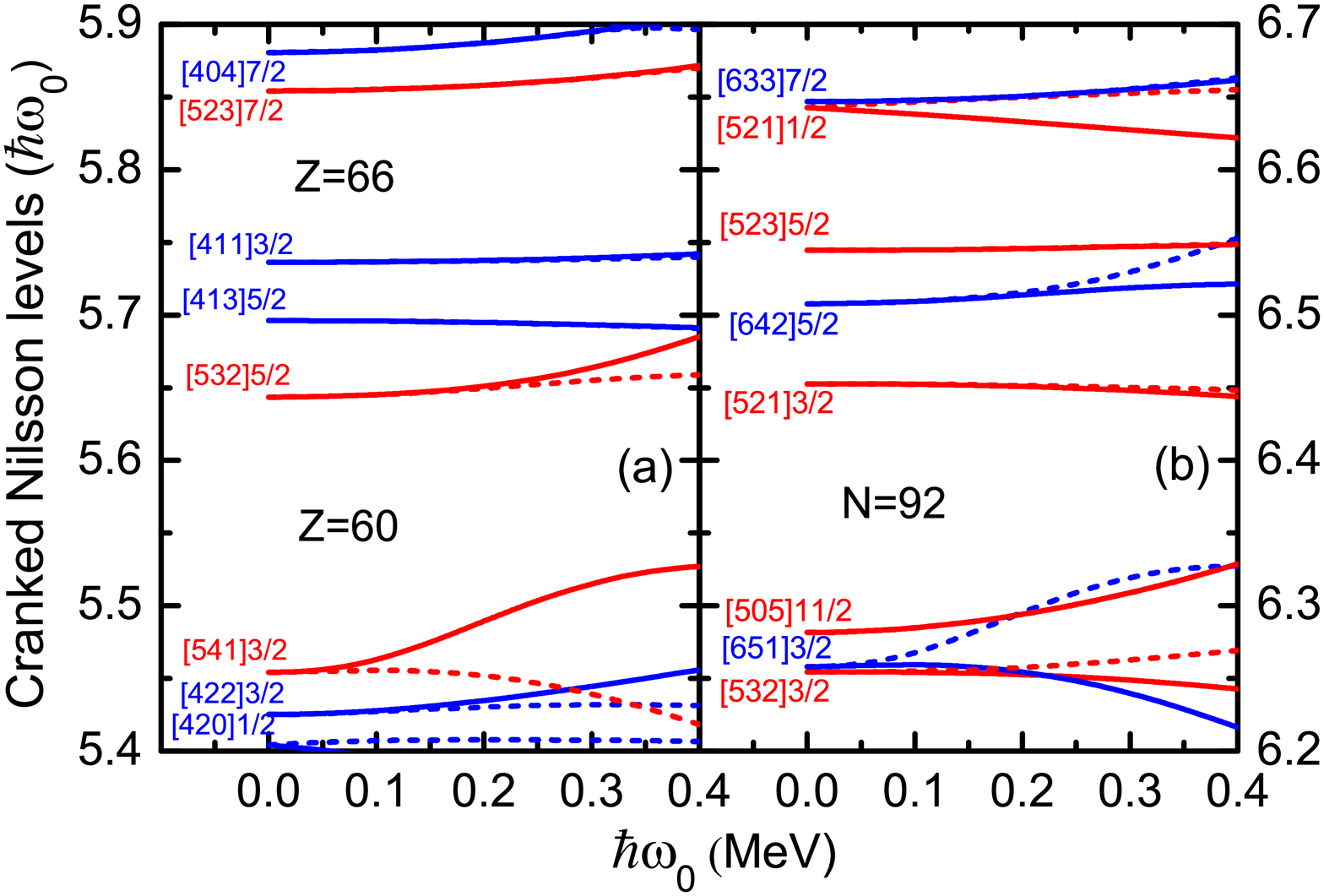}
\caption{\label{fig1:nil}
The cranked single-particle levels near the Fermi surface of $^{155}$Pm
for (a) protons and (b) neutrons.
The positive (negative) parity levels are denoted by blue (red) lines.
The signature $\alpha=+1/2$ ($\alpha=-1/2$) levels
are denoted by solid (dashed) lines.
}
\end{figure}

The proton and neutron cranked Nilsson levels near
the Fermi surface of $^{155}$Pm are shown in Fig.~\ref{fig1:nil}.
The single-particle level structures for all Pm isotopes considered
in the present work are very close to each other,
so we only show $^{155}$Pm as an example.
The present calculation shows that the ground states for
$^{153, 155, 157}$Pm are all $\pi5/2^-[532]$, which is consistent
with the experimental data~\cite{Taniguchi1996_JPSJ65-3824,
Hwang2009_PRC80-037304, Bhattacharyya2018_PRC98-044316}.
In addition, for the neighboring nuclei Nd and Sm, the experimental data
show that the ground states for $N=93$ isotones ($^{153}$Nd and $^{155}$Sm)
are $\nu3/2^-[521]$~\cite{Hwang1997_IJMPE6-331, Reich2005_NDS104-1},
which is also reproduced by the present calculation.
Therefore, the cranked Nilsson levels adopted here are quite reasonable.

\begin{figure*}[!]
\includegraphics[width=0.7\textwidth]{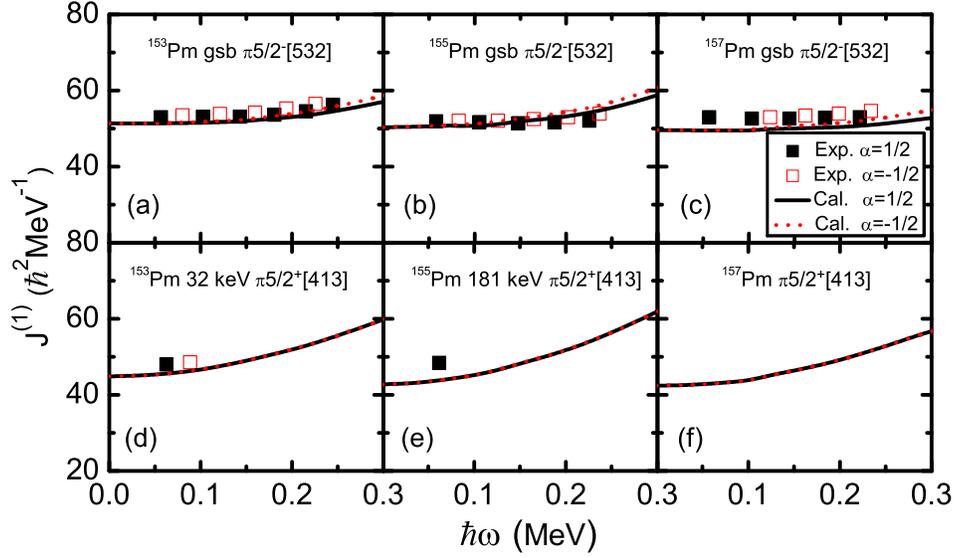}
\caption{\label{fig2:1qp}
The experimental~\cite{Bhattacharyya2018_PRC98-044316, Helmer2006_NDSs107-507,
Reich2005_NDS104-1} and calculated kinematic MOIs for the ground state
band $\pi5/2^-[532]$ (upper panel) and the excited state
band $\pi5/2^+[413]$ (lower panel) in $^{153, 155, 157}$Pm.
}
\end{figure*}

Figure~\ref{fig2:1qp} shows the comparison between
experimental~\cite{Bhattacharyya2018_PRC98-044316, Helmer2006_NDSs107-507,
Reich2005_NDS104-1} and calculated kinematic MOIs for the ground state
band (GSB) $\pi5/2^-[532]$ (upper panel) and the excited state band
$\pi5/2^+[413]$ (lower panel) in $^{153, 155, 157}$Pm.
It can be seen in Fig.~\ref{fig2:1qp} that all the experimental MOIs can be
reproduced quite well by PNC-CSM except $\pi5/2^+[413]$ in $^{155}$Pm,
which are a little underestimated by the calculation.
In addition, the signature splittings in $\pi5/2^-[532]$
are also well reproduced.
Thus, the present calculation in turn supports the configuration assignments
for these 1-qp bands in Ref.~\cite{Bhattacharyya2018_PRC98-044316}.
Note that the parameter set we adopted in the present PNC-CSM calculation
can also reproduce the rotational bands in the neighboring Nd and Sm isotopes,
including both even-even and odd-$A$ nuclei~\cite{Zhang2018_PRC98-034304}.
Therefore, PNC-CSM is reliable to make the configuration and bandhead spin assignments
for the 2-qp bands observed in odd-odd $^{154, 156}$Pm~\cite{Bhattacharyya2018_PRC98-044316}.

\begin{figure*}[!]
\includegraphics[width=0.75\textwidth]{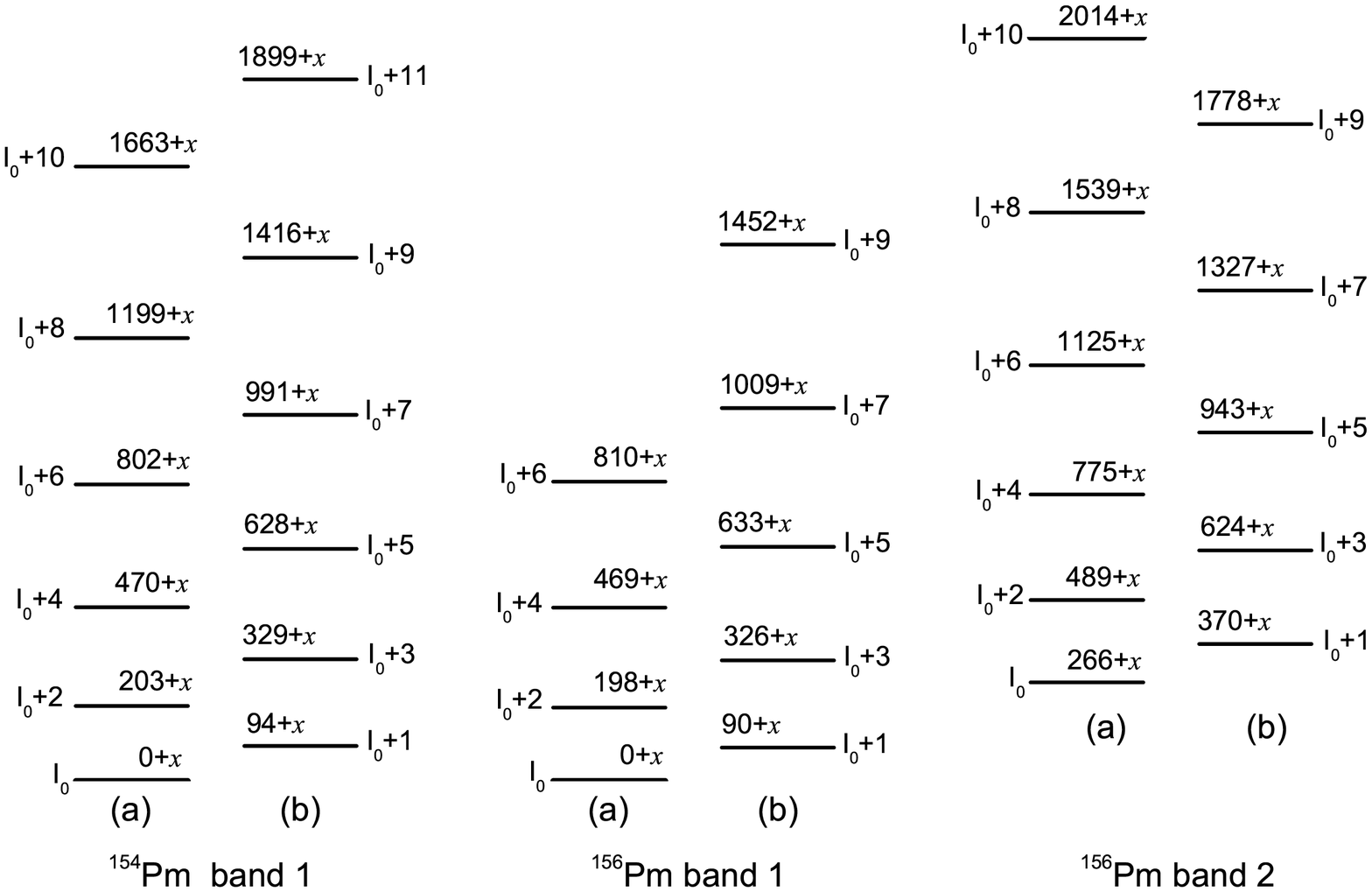}
\caption{\label{fig3:exp}
The experimental level scheme of the three 2-qp bands in $^{154}$Pm and $^{156}$Pm.
The data are taken from Ref.~\cite{Bhattacharyya2018_PRC98-044316}.
}
\end{figure*}

Figure~\ref{fig3:exp} shows the experimental level scheme of
three 2-qp bands observed in $^{154}$Pm and $^{156}$Pm.
The data are taken from Ref.~\cite{Bhattacharyya2018_PRC98-044316}.
For $^{154}$Pm, two isomers with half-lives of 2.68 minutes and 1.73 minutes
were observed several years ago~\cite{Reich2009_NDS110-2257}.
In Ref.~\cite{Sood2012_JPG39-095107}, using the quasiparticle-rotor model,
the 2.68m and 1.73m isomers have been assigned as 2-qp
with the configuration $K^\pi=4^+$ ($\pi5/2^-[532]\otimes \nu3/2^-[521]$),
and $K^\pi=1^-$ ($\pi5/2^+[413]\otimes \nu3/2^-[521]$), respectively.
In addition, the configurations for several levels have also been
assigned in Ref.~\cite{Sood2012_JPG39-095107}.
For $^{156}$Pm, the 26.7s ground state and one isomeric state with 150.3~keV
have been observed by previous experiment~\cite{Reich2003_NDS99-753}.
In Refs.~\cite{Sood2011_PRC83-027303, Sood2011_PRC83-057302},
they are interpreted as one Gallagher-Moszkowski (GM) doublets
with $K^\pi=4^+$ and $1^+$ ($\pi5/2^-[532]\otimes \nu3/2^-[521]$).
Whereas in Ref.~\cite{Hellstroem1990_PRC41-2325}, the ground state of $^{156}$Pm
is assumed to have the configuration $K^\pi=4^-$ ($\pi5/2^+[413]\otimes \nu3/2^-[521]$).
The three 2-qp rotational bands observed in Ref.~\cite{Bhattacharyya2018_PRC98-044316}
may either be established above these isomeric states,
or above another excited states which are close to these isomeric states with energy.
Due to the lack of firm spin and parity assignments, reasonable configuration assignments
have not been made for these three 2-qp bands~\cite{Bhattacharyya2018_PRC98-044316}.
Therefore, all the lowest spins in these three bands
are assumed as $I_0$ with energy $0+x$.
Note that the 2-qp band in $^{152}$Pm observed in Ref.~\cite{Bhattacharyya2018_PRC98-044316}
does not show a typical rotational character,
so we only focus on the three 2-qp bands in $^{154}$Pm and $^{156}$Pm.
In the following, the configurations and bandhead spins will be assigned for them.

\begin{table}[!]
 \centering
 \caption{\label{tab:conf}
The possible low-lying 2-qp configurations in $^{154,156}$Pm
and the corresponding $K$ quantum numbers.
The stars are added after the GM favored $K$ values.}
\begin{tabular*}{1.0\columnwidth}{c@{\extracolsep{\fill}}lll}
\hline
\hline
      & Configuration                        & $K_>$    & $K_<$    \\
\hline
Conf1 & $\pi5/2^-[532]\otimes \nu3/2^-[521]$ & $4^\ast$ & 1        \\
Conf2 & $\pi5/2^-[532]\otimes \nu5/2^+[642]$ & $5^\ast$ & 0        \\
Conf3 & $\pi5/2^-[532]\otimes \nu5/2^-[523]$ & 5        & $0^\ast$ \\
Conf4 & $\pi5/2^+[413]\otimes \nu3/2^-[521]$ & 4        & $1^\ast$ \\
Conf5 & $\pi5/2^+[413]\otimes \nu5/2^+[642]$ & 5        & $0^\ast$ \\
Conf6 & $\pi5/2^+[413]\otimes \nu5/2^-[523]$ & $5^\ast$ & 0        \\
\hline
\hline
\end{tabular*}
\end{table}

For the deformed odd-odd nucleus, when one unpaired proton and one unpaired neutron are coupled,
the projections of their total angular momentum on the symmetry axis
($\Omega_{\rm p}$ and $\Omega_{\rm n}$) can produce two states with
$K_> =|\Omega_{\rm p}+ \Omega_{\rm n}|$ and $K_< =|\Omega_{\rm p} - \Omega_{\rm n}|$
due to the residual proton-neutron interaction.
They follow the GM coupling rules~\cite{Gallagher1958_PR0111-1282}
\begin{eqnarray}\label{eq:gm}
 K_> &=& |\Omega_{\rm p} + \Omega_{\rm n}|, \
         \text{if}  \ \Omega_{\rm p}=\Lambda_{\rm p} \pm \frac{1}{2} \
         \text{and} \ \Omega_{\rm n}=\Lambda_{\rm n} \pm \frac{1}{2} \ , \nonumber\\
 K_< &=& |\Omega_{\rm p} - \Omega_{\rm n}|, \
         \text{if}  \ \Omega_{\rm p}=\Lambda_{\rm p} \pm \frac{1}{2} \
         \text{and} \ \Omega_{\rm n}=\Lambda_{\rm n} \mp \frac{1}{2} \ . \nonumber
\end{eqnarray}
Table~\ref{tab:conf} shows the possible low-lying 2-qp configurations
in $^{154,156}$Pm and the corresponding $K$ quantum numbers.
The stars are added after the GM favored $K$ values.
For convenience, these configurations are referred as Conf1 to Conf6, respectively.

It can be seen from Eq.~(\ref{eq:exp-moi}) that for one rotational band,
the extracted MOIs with rotational frequency are very sensitive to the bandhead spin.
Changing the $K$ value for one rotational band can only affect the extracted rotational frequency.
Therefore, firstly we can assign different bandhead spins and $K$ values
to these three 2-qp bands and extract the variation of the MOIs with rotational frequency.
Then by comparison with the PNC-CSM calculations using different configurations,
the configurations and bandhead spins of these 2-qp bands can be obtained.
Note that the bandhead spin of the GSB in the superheavy nucleus ${}^{256}$Rf has
already been assigned successfully by PNC-CSM using this method~\cite{Zhang2013_PRC87-054308}.
In addition, $E2$ transitions are also quite important to the configuration assignment for a rotational band.
In cranking calculations, the $E2$ transitions cannot be calculated in a quantum mechanical way.
However, the semi-classical approximations have been extensively used in describing
various novel rotations, such as the magnetic, antimagnetic, and chiral
rotations~\cite{Zhao2011_PLB699-181, Zhao2011_PRL107-122501, Zhao2012_PRC85-054310,
Meng2013_FP8-55, Zhao2015_PRC92-034319, Zhao2017_PLB773-1}.
If the configurations of these novel rotations are assigned properly,
good agreements with the data can be achieved.
Since there is no experimental $E2$ transition in these Pm isotopes,
in present work, only kinematic MOIs are adopted for the configuration and bandhead spin assignments.

\begin{figure}[!]
\includegraphics[width=0.95\columnwidth]{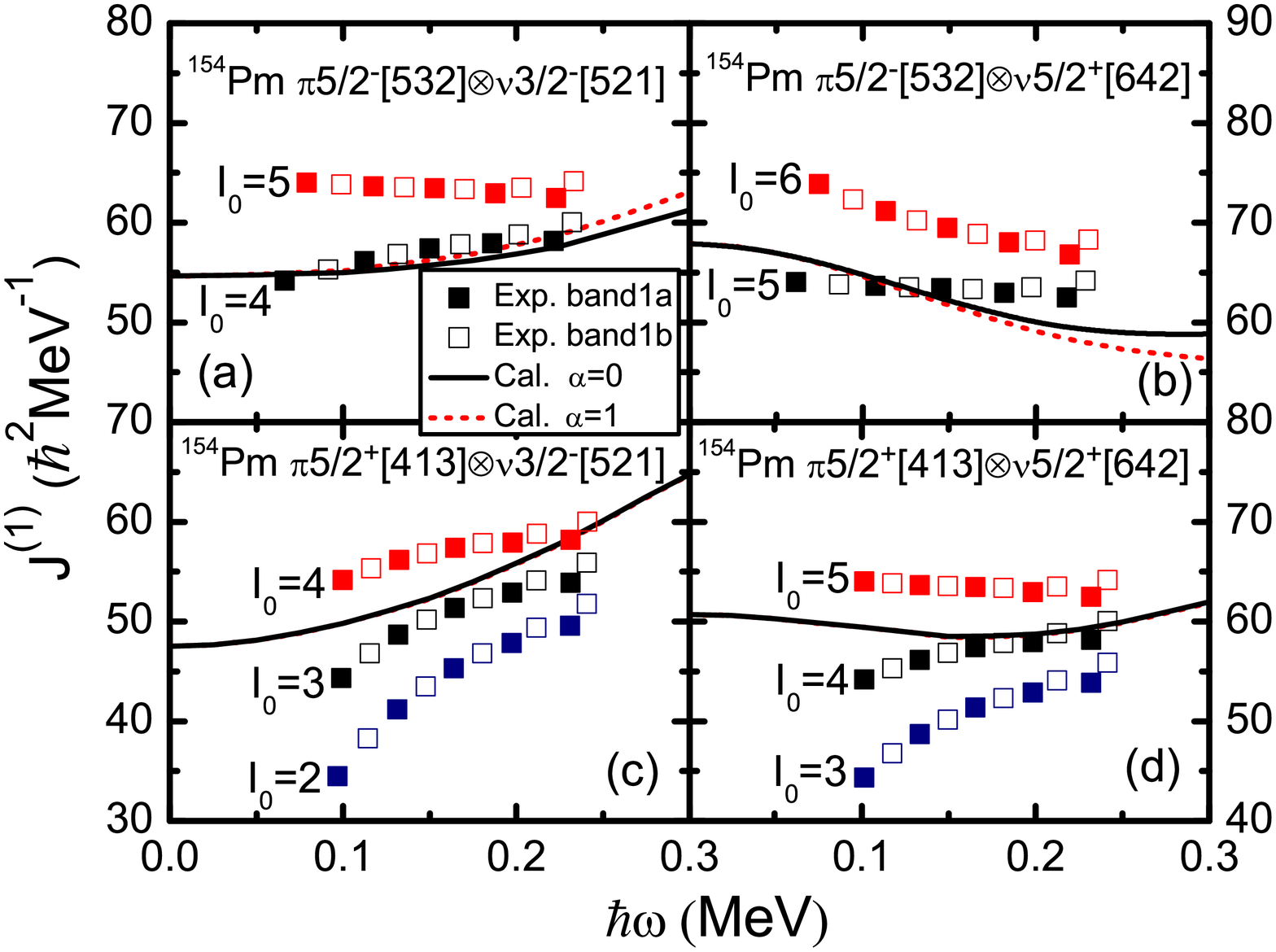}
\caption{\label{fig4:154}
The comparison between the experimental and calculated kinematic MOIs for the 2-qp
band in $^{154}$Pm (c.f., the left column of Fig.~\ref{fig3:exp})
with different bandhead spin assignments $I_0$ using the configuration
(a) $\pi5/2^-[532]\otimes \nu3/2^-[521]$,
(b) $\pi5/2^-[532]\otimes \nu5/2^+[642]$,
(c) $\pi5/2^+[413]\otimes \nu3/2^-[521]$, and
(d) $\pi5/2^+[413]\otimes \nu5/2^+[642]$.
The GM favored coupling mode (c.f., Table~\ref{tab:conf})
is chosen for each configuration assignment.
The extracted experimental MOIs far away from the
PNC-CSM calculation are not shown.
}
\end{figure}

Figure~\ref{fig4:154} shows the comparison between the experimental
and calculated kinematic MOIs for the 2-qp rotational band in $^{154}$Pm
(c.f., the left column of Fig.~\ref{fig3:exp}) with different
bandhead spin assignments $I_0$ using the configuration
(a) $\pi5/2^-[532]\otimes \nu3/2^-[521]$,
(b) $\pi5/2^-[532]\otimes \nu5/2^+[642]$,
(c) $\pi5/2^+[413]\otimes \nu3/2^-[521]$, and
(d) $\pi5/2^+[413]\otimes \nu5/2^+[642]$.
Note that the proton-neutron residual interaction is not taken into account in the PNC-CSM,
so it is hard to assign the $K$ quantum numbers for these 2-qp configurations.
Since this 2-qp rotational band may be established above the ground state of $^{154}$Pm,
the GM favored coupling mode (c.f., Table~\ref{tab:conf}) is adopted when
extracting the experimental data by Eq.~(\ref{eq:exp-moi}).
It can be seen clearly in Fig.~\ref{fig4:154} that the configuration
$\pi5/2^-[532]\otimes \nu3/2^-[521]$ [Fig.~\ref{fig4:154}(a)] can reproduce the
extracted MOIs quite well if the bandhead spin is assigned as $I_0=4\hbar$.
In addition, the experimental MOIs show a small signature splitting,
which is also reproduced quite well by the
PNC-CSM calculations [$\pi5/2^-[532](\alpha=\pm 1/2)\otimes \nu3/2^-[521] (\alpha=-1/2)$].
For other configurations [Figs.~\ref{fig4:154}(b), (c), and (d)],
no matter which bandhead spin is assigned,
all the calculations can not reproduce the extracted MOIs.
This demonstrates that the configuration of this 2-qp rotational band in $^{154}$Pm
is $\pi5/2^-[532]\otimes \nu3/2^-[521]$, and the corresponding bandhead spin is $I_0=4\hbar$.
It also can be seen from the cranked Nilsson levels in Fig.~\ref{fig1:nil} that,
this configuration corresponds to the ground state in $^{154}$Pm.
According to the GM coupling rules,
the favored $K$ value for this configuration is $K_>=4$.
Note that in Ref.~\cite{Sood2012_JPG39-095107}, the 2.68m isomer in $^{154}$Pm
is assigned as the ground state with the same configuration $K^\pi = 4^+$.

\begin{figure}[!]
\includegraphics[width=0.95\columnwidth]{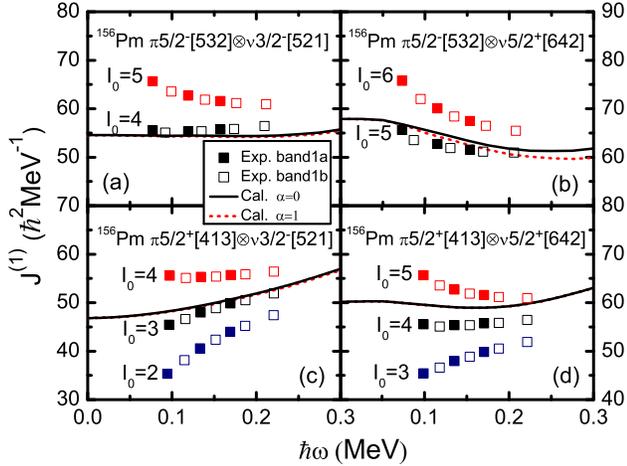}
\caption{\label{fig5:1561}
Similar with Fig.~\ref{fig4:154},
but for the band 1 in $^{156}$Pm (c.f., the middle column of Fig.~\ref{fig3:exp}).
}
\end{figure}

Figure~\ref{fig5:1561} is similar with Fig.~\ref{fig4:154}, but for the band 1
in $^{156}$Pm (c.f., the middle column of Fig.~\ref{fig3:exp}).
It can be seen that the configuration
$\pi5/2^-[532]\otimes \nu3/2^-[521]$ [Fig.~\ref{fig5:1561}(a)] can reproduce the
extracted MOIs quite well with the bandhead spin $I_0=4\hbar$.
According to the GM coupling rules,
the favored $K$ value for this configuration is $K_>=4$.
We can get that it has the same configuration as the 2-qp band in $^{154}$Pm.
In addition, this band shows nearly no signature splitting,
which is different from the 2-qp band in $^{154}$Pm.
This indicates that band 1 in $^{156}$Pm is coupled from the favored signature of
the odd proton ($\alpha = -1/2$) with $\alpha =\pm 1/2$ of the odd neutron to form
the total signature $\alpha = 0, 1$.
Note that although the configuration $\pi5/2^-[532]\otimes \nu5/2^+[642]$
seems also can approximately reproduce the extracted MOIs with bandhead spin
$I_0=5\hbar$, an obvious signature splitting exists in the calculated MOIs,
which is inconsistent with the data.
This is because both $\pi5/2^-[532]$ and $\nu5/2^+[642]$ have
signature splittings, no matter how they are coupled,
an obvious signature splitting always exists.
It also can be seen from the cranked Nilsson levels in Fig.~\ref{fig1:nil} that,
this configuration corresponds to the ground state in $^{156}$Pm.
The experimental data show that the ground states of $N=93$
isotones ($^{153}$Nd and $^{155}$Sm) are $\nu3/2^-[521]$.
From a systematic point of view, with two neutrons increasing,
the ground state of $N=95$ isotones should be $\nu5/2^+[642]$.
However, in Ref.~\cite{Hwang2008_PRC78-014309} the data show that the ground
state of $^{155}$Nd is $\nu3/2^-[521]$.
Therefore, whether this state is the ground state or not still needs further investigation.
Note that in Refs.~\cite{Sood2011_PRC83-027303, Sood2011_PRC83-057302}, the 26.7s ground
state is interpreted as $K^\pi=4^+$ ($\pi5/2^-[532]\otimes \nu3/2^-[521]$).

\begin{figure}[!]
\includegraphics[width=0.95\columnwidth]{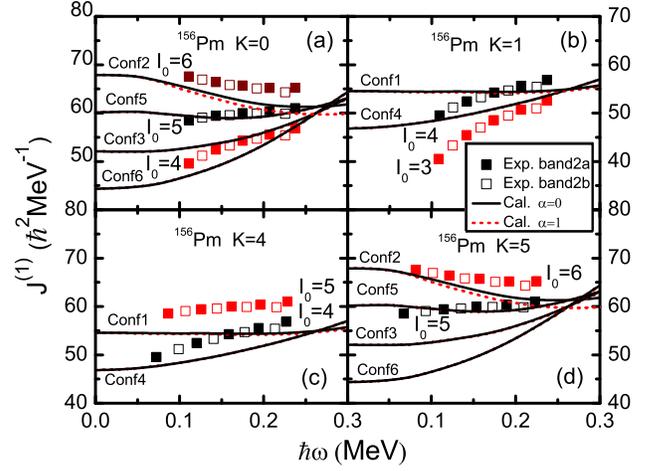}
\caption{\label{fig6:1562}
The comparison between the experimental and calculated kinematic MOIs
for the band 2 in $^{156}$Pm (c.f., the right column of Fig.~\ref{fig3:exp})
with different bandhead spin assignments using the configurations with
(a) $K=0$, (b) $K=1$, (c) $K=4$, and (d) $K=5$.
The denotations for these configurations can be seen in Table~\ref{tab:conf}.
}
\end{figure}

Since the band 2 in $^{156}$Pm (c.f., the right column of Fig.~\ref{fig3:exp})
is not established above the ground state, in Fig.~\ref{fig6:1562},
we have calculated all possible low-lying 2-qp
configurations with different bandhead spin assignments and compared with
the extracted MOIs by assigning (a) $K=0$, (b) $K=1$, (c) $K=4$, and (d) $K=5$.
The denotations for these configurations can be seen in Table~\ref{tab:conf}.
It can be seen that the configuration $\pi5/2^+[413]\otimes \nu5/2^+[642]$
[Figs.~\ref{fig6:1562} (a) and (d)]
can reproduce the extracted MOIs very well with the bandhead spin $I_0=5\hbar$,
no matter the $K$ quantum number is assigned as 5 or 0.
According to the GM coupling rules, the favored $K$ value is 0 for this configuration.
However, if $K=0$ is assigned, the bandhead should be $I=0$,
which is too far away from the assigned bandhead $I=5\hbar$.
Therefore, we tentatively assign this configuration with $K=5$.
It also can be seen that the experimental data show quite small signature splitting.
This indicates that band 2 in $^{156}$Pm is coupled from the favored signature of
the odd neutron ($\alpha = 1/2$) with $\alpha =\pm 1/2$ of the odd proton to form
the total signature $\alpha = 0, 1$.

Finally, we have made proper configuration and bandhead spin assignments
for these three 2-qp rotational bands in $^{154, 156}$Pm by PNC-CSM.
In addition, it can be seen in Figs.~\ref{fig4:154}, \ref{fig5:1561}, and~\ref{fig6:1562}
that the calculated MOIs for the 2-qp rotational bands
are quite different with different configurations.
Therefore, they may provide the information on the configuration
and bandhead spin to the rotational bands observed in further experiments.

\section{Summary}{\label{Sec:Summary}}

In summary, the recently observed rotational bands in the
neutron-rich nuclei $^{153-157}$Pm
are investigated by a particle-number conserving method.
The kinematic moments of inertia for the 1-quasiparticle bands in
$^{153,155,157}$Pm are reproduced very well by the calculation.
Configuration and bandhead spin assignments have been made
for the three 2-quasiparticle bands in $^{154, 156}$Pm
by comparison of the experimental and calculated moments of inertia.
For the 2-quasiparticle band in $^{154}$Pm, the configuration is assigned
as $\pi5/2^-[532]\otimes \nu3/2^-[521]$ ($K^\pi=4^+$)
with the bandhead spin $I_0=4\hbar$.
In $^{156}$Pm, the configurations of the two 2-quasiparticle bands are assigned
as $\pi5/2^-[532]\otimes \nu3/2^-[521]$ ($K^\pi=4^+$)
with the bandhead spin $I_0=4\hbar$,
and $\pi5/2^+[413]\otimes \nu5/2^+[642]$  ($K^\pi=5^+$)
with the bandhead spin $I_0=5\hbar$, respectively.
Meanwhile, the moments of inertia for several possible low-lying
2-quasiparticle bands in $^{154, 156}$Pm have also been calculated,
which are quite different from each other.
Therefore, these calculated results also provide valuable information on the
configuration and bandhead spin to further experiments about these two odd-odd nuclei.

\section{Acknowledgement}

This work is Supported by National Natural Science
Foundation of China (11875027, 11775112, 11775026, 11775099, 11975096),
Fundamental Research Funds for the Central Universities (2018MS058)
and the program of China Scholarships Council (No. 201850735020).


%

\end{document}